\documentclass[floatfix,twocolumn,showpacs,amsmath,amssymb,superscriptaddress,aps,prc]{revtex4-2}
\usepackage[dvips]{graphicx}
\usepackage{xcolor}
\usepackage{hyperref}
\usepackage{mathtools} % Requires amsmath
\newcommand {\cc} {\mbox{$^{12}$C+$^{12}$C}}
\newcommand{\ane}{\mbox{$\alpha+^{20}$Ne}}
\newcommand{\carb}{^{12}\mathrm{C}}
\newcommand{\neon}{^{20}\mathrm{Ne}}

\begin{document}
	\title{Towards a microscopic description of $\cc$ fusion at stellar energies}
	\author{Pierre Descouvemont}
	\email{pierre.descouvemont@ulb.be}
	\affiliation{D\'epartement de Physique, C.P. 229,
		Universit\'e Libre de Bruxelles (ULB), B 1050 Brussels, Belgium}
	\date{\today}
\begin{abstract}
I present a fully microscopic description of the $\cc$ fusion reaction at stellar energies.
Utilizing the multichannel Resonating Group Method (RGM),
my model explicitly includes $\cc$ and $\ane$ reaction channels (with excited states). This approach provides a consistent, simultaneous, description of fusion, elastic scattering, and $^{24}\mathrm{Mg}$ spectroscopy.
Results for $\cc$ elastic scattering show excellent agreement with experimental data, significantly improving the single-channel approximations.
Spectroscopic analysis reveals that $^{24}\mathrm{Mg}$ states and resonances are highly mixed configurations, contradicting the concept of pure "molecular states."
The calculated fusion $S$-factor is consistent with available experimental data and predicts both narrow and broad resonances near the Coulomb barrier.
Main resonance widths originate primarily from the $\ane$ exit channels.
The $S$-factor exhibits a decrease at low energies, providing a microscopic support for the hypothesis of fusion hindrance.
This work is a first step towards a reliable theoretical extrapolation of the $\cc$ reaction to deep stellar burning temperatures. Future works should include the neutron and proton channels to provide a complete description of the $\cc$ fusion.
\end{abstract}
	\maketitle

\section{Introduction}

The $\cc$ fusion reaction plays an important role in stellar nucleosynthesis~\cite{WIP97,BHP12}. It appears to be the main carbon-burning phase in massive stars, and triggers the production of heavy elements at high temperatures~\cite{BEJ14,WBB25,CCD25}. Determining the  $\cc$ fusion cross section at stellar energies (typically $\approx 2.4$ MeV at $T =  10^9$ K) is extremely difficult. The common problem in low-energy reactions is the smallness of the cross sections, as a consequence of the Coulomb repulsion. This problem is even more severe in heavy-ion reactions where Coulomb effects are further enhanced, compared to $(p,\gamma)$ or $(\alpha,\gamma)$ reactions.

The literature on $\cc$ fusion is extremely vast, and I refer to Refs.~\cite{WBB25,CCD25} for recent reviews. The  $\cc$ reaction presents two specificities which make extrapolation very uncertain. The presence of resonances in $\cc$ scattering is well established for decades \cite{BKA60}. The fusion data around the Gamow energy also present resonances \cite{JBE13}, the origin of which remains unclear. A recent experiment, using the Trojan Horse method~\cite{TSL18}, claimed to find several resonances at low energies, which would have significant implications on massive-star evolution. This result, however, was subsequently challenged in Ref.~\cite{MPK19,BMT20} where the authors reanalyzed the data with an improved theoretical model. Resonances at low energies were also suggested by Spillane {\sl et al.}~\cite{SRR07}, but the large error bars do not permit a definite conclusion.

Another issue under debate is the existence (or absence) of a  $\cc$ fusion hindrance at astrophysical energies. This phenomenon was first observed for heavier systems~\cite{JER02}. Lighter systems such as  $\cc$ were then analyzed by Jiang {\sl et al.}~\cite{JRB07} who concluded that fusion hindrance may also occur in reactions such as  $\cc$. This effect would have important consequences in nuclear astrophysics~\cite{CSB09}. The current data, however, do not allow to conclude on the hindrance
effect since they do not cover energies sufficiently low. A
recent work of Uzawa and Hagino~\cite{UH25} investigates the hindrance
possibility by fitting the experimental cross sections with more flexibility.

Both physical issues discussed above, the presence of $\cc$ resonances
at low energies and the fusion hindrance, are difficult to address
by theory. Most calculations are performed within the optical
model and its multi-channel extensions \cite{GAA05,AD13,CKC18,GCC22}. In this approach, the fusion process is either simulated by an imaginary part in 
the optical potential \cite{AD13,GCC22}, or determined through a barrier penetration model of quantum tunneling \cite{GAA05,CKC18}.  The standard
methodology is to fit the available data, and to extrapolate the theory
down to stellar energies. The optical model, however, is unable to reproduce bound or resonant states of
the system. Although they do not play a direct role in fusion reactions,
these states provide useful tests of the model.
Other approaches, such as the time-dependent Hartree-Fock method \cite{SU18,CSD25} or the time-dependent wave packet method \cite{DW18}, have been developed. These techniques, however, are limited to the fusion component of the $\cc$ reaction, and cannot be applied to the $^{24}$Mg spectroscopy. 
 
These difficulties can be addressed by microscopic models, where the nucleon-nucleon interaction is the only input~\cite{WT77}. One of the main advantages of microscopic models is to provide a simultaneous description of fusion, of elastic scattering and of the $^{24}$Mg spectroscopy. However,
solving a many-body problem (i.e., $A \gtrsim 6$) is an extremely difficult
task, especially for scattering problems. Recent advances in \emph{ab initio} methods~\cite{He20,QKN25} consider $\alpha +p$~\cite{QN09} or $\alpha + d$~\cite{NQ11} scattering, but extending these methods to heavier nuclei is currently not possible. 

An efficient technique to overcome this limitation is the cluster approximation~\cite{WT77}. In the Resonating Group Method (RGM, see Ref.~\cite{Ho77}), the wave function is obtained from cluster wave functions defined in the
shell model. Early calculations focused on simple systems,
such as $\alpha+p$, but the development of computer capabilities
allowed to go far beyond this limit (see, for example, Ref.~\cite{TD10}).

A microscopic approach to fusion has been considered previously. In Ref.~\cite{De89b}
a single-channel model of $\cc$ was used, with a phenomenological
imaginary part fitted to the data. In Ref.~\cite{TK24} the authors used the
Antisymmetrized Molecular Dynamics (AMD) model, to investigate the
$^{24}\text{Mg}$ nucleus above the $\cc$ threshold. This approach, however, is based on
the bound-state approximation. The authors determine the widths of resonances from the reduced-width amplitudes, and use the Breit-Wigner approximation for each resonance. This method, however, does not include scattering conditions, and neglects the interference between the resonances.

In the present work, I perform a fully microscopic calculation of $\cc$
fusion, by including the dominant $\ane$ channels ($\neon$ in the ground state and in several excited states). With the inclusion of $\alpha$ channels, the main component of the fusion cross section is explicitly taken into account. No phenomenological imaginary potential is necessary. The $\carb$ and $\neon$
clusters are described in the shell model, by considering all $p$-shell and
$sd$-shell configurations, respectively. This accurate description leads to very large
bases: $\cc$ involves $225^2=50625$ Slater determinants. A large number
of $^{12}\text{C}$ and $^{20}\text{Ne}$ states are included in a consistent way. This approach simultaneously provides $^{24}\text{Mg}$ properties and $\cc$ elastic
scattering and fusion. In this exploratory study, I omit the $^{23}\text{Na}+p$
and $^{23}\text{Mg}+n$ configurations which involve even larger numbers of Slater determinants,
and are beyond the scope of the present work.

The paper is organized as follows. In Sec.~\ref{sec2}, I present the microscopic
cluster model, and the calculation of the fusion cross section. Section~\ref{sec3} is
dedicated to $^{12}\text{C}$ and $^{20}\text{Ne}$ in the shell model, and Sec.~\ref{sec4} to the 24-nucleon wave functions. I present various tests of
the model with $^{24}\text{Mg}$ spectroscopy and $\cc$ elastic scattering in Sec.~\ref{sec5}. I also discuss the role of the inelastic and reaction channels. In
Sec.~\ref{sec6}, I discuss more specifically the fusion cross section and the role of resonances. Conclusions
and perspectives are presented in Sec.~\ref{sec7}.
	
\section{The Microscopic Model}
\label{sec2}
	
My goal is to describe the $\cc$ fusion reaction within a fully microscopic approach. This means that the cross sections are derived from a nucleon-nucleon interaction, and that the asymptotic boundary conditions are taken into account explicitly. The $A$-nucleon Hamiltonian ($A=24$ here) is given by
	\begin{align}
	H = \sum_{\alpha=1}^{A} t_{\alpha} + \sum_{\alpha <\beta=1}^{A} v_{\alpha \beta},
	\label{eq1}
	\end{align}
where $t_{\alpha}$ is the kinetic energy of nucleon ${\alpha}$, and $v_{\alpha \beta}$ is a nucleon-nucleon interaction [notice that three-body terms are neglected in Eq.~\eqref{eq1}].
	
Solving the Schr\"{o}dinger equation associated with \eqref{eq1} is an extremely complicated problem, which can only be treated approximately. In the present work, I use the multichannel Resonating Group Method (RGM, see Ref.~\cite{WT77}) which is equivalent to the Generator Coordinate Method (GCM, see Ref.~\cite{Ho77}). In addition to $\cc$ channels, some $\ane$ channels are open at the threshold, and play a crucial role in the fusion process. They must be included in the model to obtain a realistic description of the fusion process. 

Schematically, the RGM wave functions for angular momentum $J$ are therefore given by two components written as
	\begin{align}
	\Psi^J = \Psi_{\rm{C+C}}^J + \Psi_{\alpha+\rm{Ne}}^J,
	\label{eq2}
	\end{align}
which are associated with the $\cc$ and $\ane$ partitions, respectively (notice that I only consider positive-parity states owing to the symmetry of the entrance channel). The RGM is based on the cluster approximation which provides, for the symmetrized  $\cc$ component,
	\begin{align}
	\Psi_{\rm{C+C}}^J = \mathcal{A} \sum_{ij} g_{ij}^J(\rho) \bigl(\Phi_{{\rm C}(i)}  \Phi_{{\rm C}(j)}+\Phi_{{\rm C}(j)}  \Phi_{{\rm C}(i)}\bigr)/(1+\delta_{ij}),
	\label{eq3}
	\end{align}
where $\mathcal{A}$ is the $A$-nucleon antisymmetrizer, $\Phi_{{\rm C}(i)}$ are $\carb$ wave functions, defined in the shell model, and $g_{ij}^J(\rho)$ are radial  functions depending on the $\cc$ relative coordinate $\rho$. The summation over indices $i$ and $j$ corresponds to different $^{12}$C states.

For the sake of simplicity, I neglect angular momentum couplings in the definition \eqref{eq3}. 
In practice, the spins of both $\carb$ nuclei are coupled to form the channel spin $I$, which in turn is coupled to the relative orbital momentum $\ell$. Channels with $i=j$ are symmetrized by keeping only even values of $I$.
	
Similarly, the second component in Eq.~\eqref{eq2} is defined by
	\begin{align}
	\Psi_{\alpha+{\rm Ne}}^J =\mathcal{A} \sum_k \tilde{g}_k^J(\tilde{\rho}) \Phi_\alpha \Phi_{{\rm Ne}(k)},
	\label{eq4}
	\end{align}
where $\Phi_\alpha$ is a $(0s)^4$ shell-model wave function of the $\alpha$ particle, and where $ \Phi_{\rm{Ne}(k)}$ are $^{20}$Ne wave functions, defined with 2 neutrons and 2 protons in the $sd$ shell-model space. The relative functions $\tilde{g}_k^J(\tilde{\rho})$, depending on the $\ane$ relative coordinate $\tilde{\rho}$ are determined from the Schr\"{o}dinger equation in the same way as in Eq.~\eqref{eq3}.
	
A decisive advantage of cluster models is that the RGM wave functions \eqref{eq3} and \eqref{eq4} are consistently adapted to scattering problems, and to spectroscopy. The various cross sections (elastic and inelastic scattering, fusion, etc.) are defined from the asymptotic part of the radial wave functions. Assuming that the entrance channel is 
$^{12}\rm{C}(g.s.) +^{12}\rm{C}(g.s.)$, referred to as "channel 1", the asymptotic form of the scattering wave function reads
	\begin{align}
		g^J_1(\rho) \rightarrow & \sum_{ij} \Bigl[ I_J(k_1 \rho)\delta_{i1}\delta_{j1} 
		 - U_{1,ij}^J O_J(k_{ij}\rho) \Bigr] \Phi_{{\rm C}(i)}  \Phi_{{\rm C}(j)}\nonumber \\
	&	- \sum_k  U_{1,k}^J O_J(\tilde{k}_k \tilde{\rho}) \Phi_\alpha \Phi_{\rm{Ne}(k)}.
	\label{eq5}
	\end{align}
In this definition, $I_J(x)$ and $O_J(x)$ are the incoming and outgoing Coulomb functions, and $k_{ij}$ and $\tilde{k}_k$ are the wave numbers in the $\cc$ and $\ane$ channels.
Note that definition \eqref{eq5} assumes that all channels are open, which is not true at low energies. However, the extension is fairly simple: for closed channels, the Coulomb functions are replaced by appropriate Whittaker functions. As mentioned before, definition \eqref{eq5} is schematic, for the sake of clarity. In actual calculations, the cluster spins are coupled to each other, and the resulting channel spin is coupled to the relative orbital momentum \cite{DD12}.
	
The important quantity in reaction calculations is the scattering matrix $\pmb{U}^J$, which appears in Eq.~\eqref{eq5}. Elements $U_{1,1}^J$ are associated with elastic scattering and are used with standard formulae to determine the elastic cross sections. Inelastic scattering ($i\neq 1$ and/or $j\neq 1$) is provided by $U_{1,ij}^J$. Elements $U_{1,k}^J$ which appear in the second term of Eq.~\eqref{eq5} are associated with the transfer to the $\ane$ channels, i.e.\ to the fusion process. 

Ideally, the RGM wave function Eq.~\eqref{eq2} should contain $p + ^{23}$Na and $n + ^{23}$Mg channels, which also play a role in the fusion process. However, owing to the extremely large number of Slater determinants ($C^{12}_4 \times C^{12}_3=108900$ for $^{23}$Na and $^{23}$Mg), introducing these channels is beyond the scope of this work. I therefore compute the $\alpha$ component of the fusion cross section, which is defined as
	\begin{align}
	\sigma_F = \frac{2\pi}{k_{\mathrm{c.m.}}^2} \sum_{J\ {\rm even}}  (2J+1) \sum _k |U_{1,k}^J|^2,
	\label{eq6}
	\end{align}
where $k_{\mathrm{c.m.}}$ is the wave number in the $\cc$ entrance channel. The factor 2 and the limitation to even $J$ values stem from the symmetry of the entrance channel.
	
Microscopic cluster calculations of the $\cc$ system have been performed for many years, but are limited to a single-channel approximation and do not consider fusion \cite{BP82,BD84,SH82}. My work goes far beyond these limitations: it involves all $p$-shell states of $\carb$, and paves the way to include more channels since the main inelastic and transfer channels ($\ane$) are included. In the next section, I discuss the numerical issues of the method and show that these calculations are computationally challenging.
	
\section{$\carb$ and $\neon$ shell-model wave functions}
\label{sec3}
	
The cluster wave functions mentioned in Eqs.~\eqref{eq3} and \eqref{eq4} are defined in the shell model. Let us first discuss the $\carb$ wave functions which are based on Slater determinants $\varPhi_i$. Considering all states in the $p$ shell, I have $N=C^{6}_2 \times C^{6}_2=225$ Slater determinants $\varPhi_i$. As they do not have good quantum numbers, the diagonalization of symmetry operators (total spin $I^2$ and $I_z$, intrinsic spin $S^2$, orbital angular momentum $L^2$, and isospin $T^2$) provides basis functions defined by
	\begin{align}
	\Phi_{LSTn}^{IM} = \sum_{k=1}^{N} d_{LSTn,k}^{IM} \varPhi_k,
	\label{eq7}
	\end{align}
where $n$ is an additional quantum number associated with the degeneracy. Coefficients $d_{c,i}^{JM}$ [where index $c$ stands for $c = (L,S,T,n)$] do not depend on the Hamiltonian. The $\carb$ wave functions in \eqref{eq3} are therefore linear combinations of functions \eqref{eq7}, and read
	\begin{align}
	\Phi_{{\rm C}(i)}^{IM} = \sum_{c} f_{i,c}^I \Phi_{c}^{IM},
	\label{eq8}
	\end{align}
where coefficients $f_{i,c}^I$ are obtained from the diagonalization of the Hamiltonian in basis \eqref{eq7} (the nucleon-nucleon interaction will be specified in Sec.~\ref{sec5}). In this equation, $i=1$ corresponds to the lowest state for a given spin $I$. The ground state is $I=0$, and the first $2^+$ and $4^+$ states are also included. For each spin $I$, states with $i>1$ are referred to as pseudo-states. Although they do not correspond to physical states, they improve the description of the total wave functions. 

\begin{figure}[htb]
	\centering
	\includegraphics[width=7.6cm]{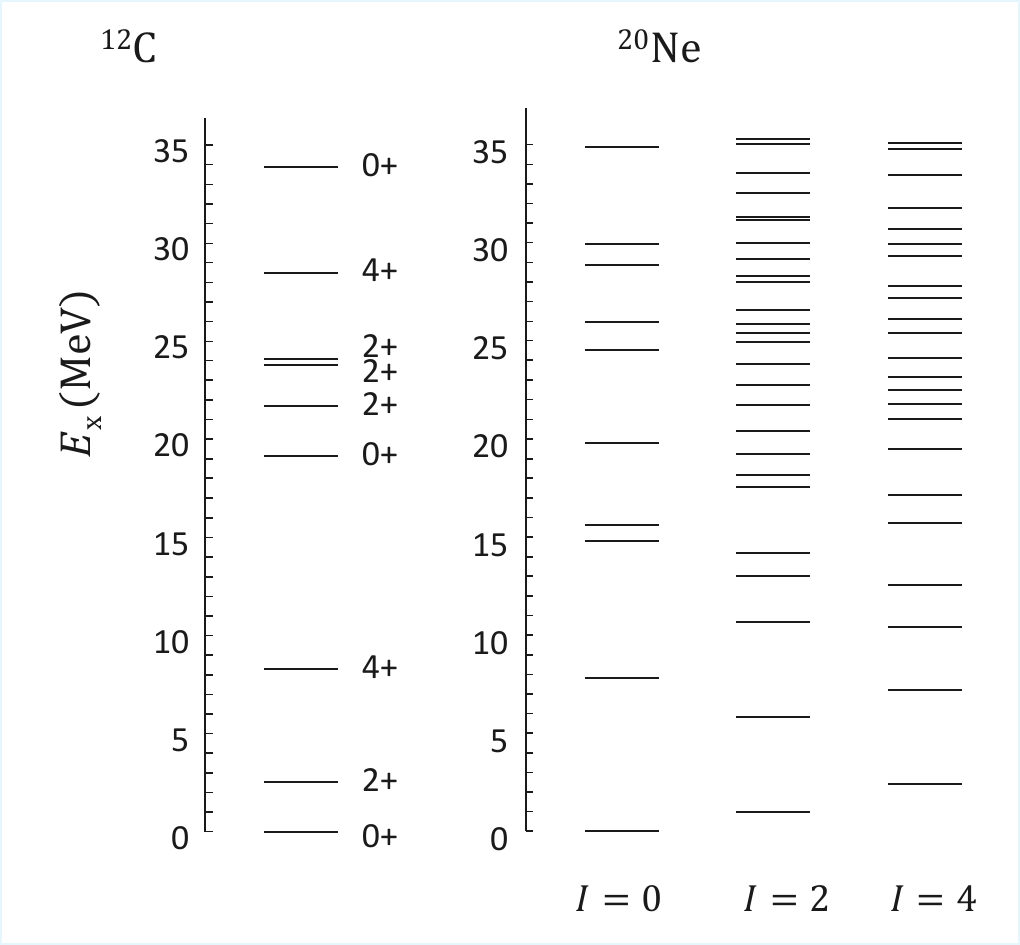}
	\caption{Shell-model $^{12}$C and $^{20}$Ne states up to 35 MeV. The Volkov V2 interaction is used (see Sec.\ \ref{sec5}).}
	\label{fig_spec1}
\end{figure}

All shell-model $\carb$ states are displayed in Fig.~\ref{fig_spec1} for energies lower than 35 MeV. The values of $(ISLTn)$ are given in Table \ref{table1}. Notice that, for the sake of simplicity, I have neglected the components with $T > 0$ or $S > 1$. From this table, I deduce that four $0^+$ states, six $2^+$ states and two $4^+$ states are included for $\carb$. Coefficients $\vert f_{i,c}^I\vert^2$ for $i=1$ are given in Table \ref{table1}. In all cases the $S=0$ component is dominant, but a small $S=1$ admixture, due to the spin-orbit force, is present. For the $4^+$ state of $\carb$, $(S=1,L=4)$ and $(S=1,L=5)$ components cannot be described in the $p$ shell.

\begin{table}[h]
	\caption{Quantum numbers $I,S,L,n$ of the $^{12}$C and $^{20}$Ne states included in the GCM wave functions. Components with $T > 0$ or $S>1$ are neglected. The last column shows the channels amplitudes in the lowest state (the contributions of degenerated states have been summed).}
	\begin{ruledtabular}
		\begin{tabular}{cccccccccc}
			$^{12}$C &  &  &  &  & $^{20}$Ne &  &  &  & \\
			$I$ & $S$ & $L$ & $n$ & $\vert f_{1,c}^I\vert^2$ & $I$ & $S$ & $L$ & $n$ &$\vert f_{1,c}^I\vert^2$ \\
			\hline
			$0$ & $0$ & $0$ & $2$ & $   0.840$ & $0$ & $0$ & $0$ & $7$ & $   0.820$\\
			& $1$ & $1$ & $2$ & $   0.160$ &  & $1$ & $1$ & $9$ & $   0.180$\\
			$2$ & $0$ & $2$ & $2$ & $   0.890$ & $2$ & $0$ & $2$ & $11$ & $   0.820$\\
			& $1$ & $1$ & $2$ & $   0.020$ &  & $1$ & $1$ & $9$ & $   0.050$\\
			& $1$ & $2$ & $1$ & $   0.060$ &  & $1$ & $2$ & $10$ & $   0.090$\\
			& $1$ & $3$ & $1$ & $   0.030$ &  & $1$ & $3$ & $11$ & $   0.040$\\
			$4$ & $0$ & $4$ & $1$ & $   0.950$ & $4$ & $0$ & $4$ & $8$ & $   0.810$\\
			& $1$ & $3$ & $1$ & $   0.050$ &  & $1$ & $3$ & $11$ & $   0.090$\\
			&  &  &  &  &  & $1$ & $4$ & $7$ & $   0.080$\\
			&  &  &  &  &  & $1$ & $5$ & $5$ & $   0.020$
		\end{tabular}
	\end{ruledtabular}
	\label{table1}
\end{table}
	
The same discussion is applied to $\neon$, which is defined with two neutrons and two protons in the $sd$ shell, the $p$ shell being complete. In this case, the number of Slater determinants increases to $N = 66^2 = 4356$. For the $\neon$ nucleus, I have sixteen $0^+$ states forty-one $2^+$ states and thirty-one $4^+$ states. The various states are displayed in Fig.~\ref{fig_spec1} and the components of the lowest states are given in Table \ref{table1}. When all angular-momentum couplings are introduced in Eqs.~\eqref{eq3} and \eqref{eq4}, the number of basis functions becomes extremely large.  For example, $J = 0^+$ contains 150 channels in the $\cc$ partition, and 88 channels in the $\ane$ partition. These numbers increase to 500 and 232 for $J = 2^+$.

\section{GCM wave functions}
\label{sec4}
\subsection{GCM kernels}
The $\carb$ and $\neon$ shell-model wave functions are then used to describe the $24$-nucleon system. In the GCM, the total $\cc$  wave function \eqref{eq2} is expressed as a combination of Slater determinants involving cluster wave functions as
	\begin{align}
	\Phi_{ij}(\pmb{R}) = \mathcal{A} 
	\varPhi_i \biggl(-\frac{\pmb{R}}{2} \biggr)
	\varPhi_j \biggl(\frac{\pmb{R}}{2}  \biggr),
	\label{eq9}
	\end{align}
where $\pmb{R}$ is the generator coordinate. The indices $i$ and $j$ run from 1 to 225, making the basis size $225^2 = 50625$ Slater determinants. Notice that I use a common oscillator parameter ($b=1.6$ fm) to exactly factorize the center-of-mass motion.

The main part of the numerical calculation is to compute the Hamiltonian and overlap kernels between basis states \eqref{eq9},
	\begin{align}
		H_{ij, kl}(\pmb{R}, \pmb{R}') &= \langle \Phi_{ij}(\pmb{R}) | H | \Phi_{kl}(\pmb{R}') \rangle \nonumber \\
		N_{ij, kl}(\pmb{R}, \pmb{R}') &= \langle \Phi_{ij}(\pmb{R}) | \Phi_{kl}(\pmb{R}') \rangle ,
	\label{eq10}
	\end{align}
which is computationally demanding ($50625^2$ matrix elements). In practice, these matrices are split into pieces, allowing for efficient parallelization of the code. The matrix elements are computed with standard formulae \cite{Br66}. The linear combinations defined in Eqs.~(\ref{eq7},\ref{eq8}) are subsequently performed. The same procedure is applied to the $\ane$ channel, and to the couplings between these channels. 
	
The next step in the calculation is to perform the angular momentum projection mentioned in Sec.~\ref{sec2}. The angular-momentum projection is carried out from integrals over the angle between $\pmb{R}$ and $\pmb{R}'$ (see Ref.~\cite{DD12} for detail). 

\subsection{Relative functions}
The relative functions in \eqref{eq3} and \eqref{eq4} can be expanded as a superposition of several generator coordinates as
	\begin{align}
	g_{ij}^{J}(\rho) \approx \sum_n F_{ij}^J(R_n) \Gamma(\rho, R_n),
	\label{eq11}
	\end{align}
where $\Gamma(\rho, R)$ are a Gaussian functions centered at $R$, and where $F_{ij}^J(R)$ is the generator function (see Refs.\ \cite{Ho77,DD12} for details). A similar expansion is applied to the $\ane$ partition.
	
The GCM technique is well suited to numerical calculations since it involves Slater determinants. However, the long-range behaviour of expansion \eqref{eq11} is Gaussian and does not correspond to the physical Coulomb function \eqref{eq5}. The problem is solved with the microscopic $R$-matrix method \cite{BHL77,DB10} which provides the scattering matrices $\pmb{U}^J$ as well as the generator functions $F_{ij}^J(R)$.

The radial wave functions $g^J_{\gamma}(\rho)$ are defined with the antisymmetrizer operator $\mathcal{A}$, defined in Eqs.~(\ref{eq3},\ref{eq4}) [index $\gamma$ stands for  $\gamma = (i,j)$]. These functions contain the so-called forbidden states \cite{Sa69}, and cannot be interpreted without the operator $\mathcal{A}$. It is, however, possible to transform these relative functions in order to derive approximations neglecting the antisymmetrizer \cite{Ho77,VL88}.

The method is based on the RGM overlap kernel, defined as
	\begin{align}
		{\mathcal N}_{\gamma,\gamma'}(\rho,\rho')=
		\langle \delta(\rho-r)\Phi_i \Phi_j\vert \mathcal{A}
			\vert \delta(\rho-r')\Phi_k \Phi_l \rangle,
	\label{eq12}
\end{align}
where $\Phi_i,\Phi_j,\Phi_k,\Phi_l$ are $^{12}\mathrm{C}$, $\alpha$ or $^{20}\mathrm{Ne}$ internal wave functions. As shown in Ref.~\cite{VL88}, the RGM and GCM kernels \eqref{eq12} and \eqref{eq10} are connected to each other by integral transforms. 

From the overlap kernel \eqref{eq12}, I define
	\begin{align}
&{\hat g}_{\gamma}(\rho) = \sum_{\gamma'} \int	{\mathcal N}^{1/2}_{\gamma,\gamma'}(\rho,\rho') g_{\gamma'}(\rho') d\rho', \nonumber \\
&{\tilde g}_{\gamma}(\rho) = \sum_{\gamma'} \int {\mathcal N}_{\gamma,\gamma'}(\rho,\rho') g_{\gamma'}(\rho') d\rho',
	\label{eq13}
\end{align}
which are normalized as
	\begin{align}
\sum_{\gamma'} {\mathcal N}_{\gamma} = 1, \quad  {\mathcal N}_{\gamma} = \int ({\hat g}_{\gamma}(\rho))^2 d\rho, 
	\label{eq14}
\end{align}
and
	\begin{align}
\sum_{\gamma} {\mathcal S}_{\gamma} = {\mathcal S}, \quad {\mathcal S}_{\gamma} = \int ({\tilde g}_{\gamma}(\rho))^2d\rho.
	\label{eq15}
\end{align}
In Eq.~\eqref{eq15}, ${\mathcal S}_{\gamma}$ is the spectroscopic factor in channel $\gamma$. Functions ${\tilde g}_{\gamma}(\rho)$ are also referred to as the reduced-width amplitudes. The total spectroscopic factor ${\mathcal S}$ is different from unity, owing to the presence of the antisymmetrizer. The numbers ${\mathcal N}_{\gamma}$ and ${\mathcal S}_{\gamma}$ provide information on the structure of a state. At large distance, functions ${\hat g}_{\gamma}(\rho)$ and ${\tilde g}_{\gamma}(\rho)$ are identical since antisymmetrization effects are negligible. Additional information about these transformations can be found in Refs.~\cite{VL88,De23}.

\section{Tests of the model: $^{24}\text{Mg}$ spectroscopy and $\cc$ elastic scattering }
\label{sec5}
\subsection{Conditions of the calculation}

The microscopic cluster calculation is based on an effective  
nucleon-nucleon interaction $v_{\alpha \beta}$ [see Eq.~\eqref{eq1}]. An important constraint   
on the fusion cross section is to reproduce the threshold energy of  
the $\ane$ channel ($Q = 4.63$ MeV). I use the Volkov V2 interaction  
\cite{Vo65} complemented by Bartlett and Heisenberg terms (with $b = -h$).
This modified interaction provides more flexibility than the original  
V2 force since it involves an additional parameter \cite{DD05b}. The interaction  
contains the exact Coulomb force and a zero-range spin-orbit term  
(with a commonly used amplitude $S_0 = 30\ \text{MeV} . \text{fm}^5$). The oscillator parameter, common to all clusters, is 1.6 fm. The generator coordinate values $R_n$ range from 1.5 fm to 10.5 fm with a step of 1 fm.

The strategy to determine the adjustable parameters (Majorana $m$, and  
Bartlett $b$) is to optimize the ground-state energy while keeping the  
threshold at the experimental value. With the multichannel model,  
$b=0$ and $m=0.70$ provide a binding energy of 11.9 MeV with respect to the $\cc$ threshold, which is  
slightly lower than the experimental value (13.9 MeV). This underestimation   
is acceptable due to the absence of \( ^{23}\text{Na} + p \) and \( ^{23}\text{Mg} + n \)  
channels.

\subsection{$^{24}$Mg bound states and resonances}
\begin{figure}[htb]
	\centering
	\includegraphics[width=8.5cm]{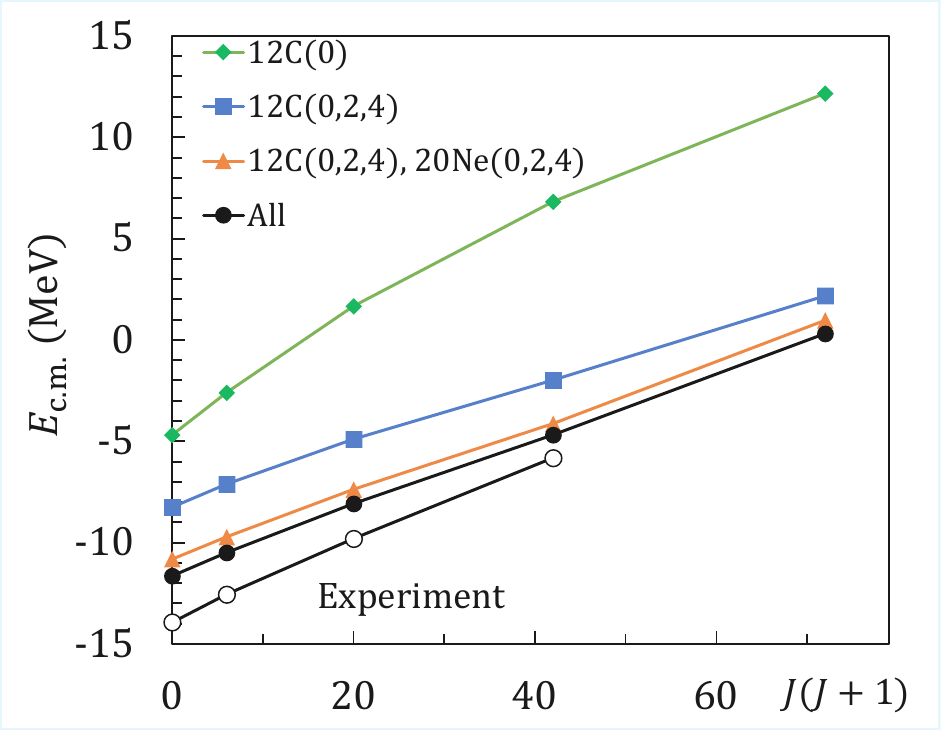}
	\caption{$^{24}$Mg lowest states for increasing number of $^{12}$C and $^{20}$Ne basis states. The zero energy is the $\cc$ threshold.}
	\label{fig_24mg}
\end{figure}

An important advantage of a microscopic approach is the  
simultaneous description of bound, resonant and continuum states of  
$^{24}\text{Mg}$. The analysis of the energy spectrum provides a test  
of the accuracy of the model; in particular, the role of $\carb$ excited  
states and of $\ane$ channels can be analyzed. In Fig.~\ref{fig_24mg}, I show  
the lowest levels of $^{24}\text{Mg} $ by introducing more and more channels. The  
single-channel approximation [labeled as "12C(0)"] provides energies far above the full calculation.  
Introducing the first $2^+$ and $4^+$ states of $\carb$ [line labeled as "12C(0,2,4)"] significantly increases the binding energies. With the $\ane$ channels, the energies are similar to the converged energies, where all channels involving pseudo-states are included (see Fig.~\ref{fig_spec1}).

The structure of the lowest $0^+_1,2^+_1$ and $4^+_1$ states is analyzed in Table \ref{table2}. I provide the amplitudes ${\cal N}_{\gamma}$ and the spectroscopic factors ${\cal S}_{\gamma}$ in some selected channels. These results suggest that the $^{24}$Mg wave functions are spread over various channels. The 	$\text{C}(0^+_1)+\text{C}(0^+_1)$ channel represents at most $10\%$ of the wave function. As shown in Fig.~\ref{fig_24mg}, the $\ane$ channels play a significant role in the spectroscopy of the low-lying $^{24}$Mg states.

\begin{table}[h]
	\caption{Spectroscopic amplitudes  ${\mathcal N}_{\gamma}$ and ${\mathcal S}_{\gamma}$ for the lowest $J=0^+,2^+,4^+$ states of $^{24}\text{Mg}$. The values are given for some typical channels and are multiplied by 100.}
\label{table2}
\begin{ruledtabular}
	\begin{tabular}{lcccccc}
Channel	$\gamma$	& ${\cal N}(0^+)$ & ${\cal S}(0^+)$ & ${\cal N}(2^+)$ & ${\cal S}(2^+)$ & ${\cal N}(4^+)$ & ${\cal S}(4^+)$\\
	$\text{C}(0^+_1)+\text{C}(0^+_1)$ & $    12.0$ & $     5.6$ & $     9.7$ & $     4.6$ & $     1.8$ & $     1.0$\\
	$\text{C}(2^+_1)+\text{C}(0^+_1)$ & $    17.9$ & $     7.5$ & $    15.7$ & $     7.2$ & $    12.4$ & $     5.9$\\
	$\text{C}(2^+_1)+\text{C}(2^+_1)$ & $     0.4$ & $     0.1$ & $     0.4$ & $     0.1$ & $     3.7$ & $     3.2$\\
	$\alpha+\text{Ne}(0^+_1)$ & $    18.9$ & $    13.9$ & $     9.8$ & $     8.0$ & $     3.4$ & $     3.1$\\
		$\alpha+\text{Ne}(2^+_1)$ & $     1.8$ & $     0.6$ & $     9.4$ & $     6.4$ & $    12.0$ & $     8.5$\\
		$\alpha+\text{Ne}(4^+_1)$ & $    12.7$ & $     8.6$ & $    13.1$ & $     9.2$ & $    16.9$ & $    12.0$\\
		All & $100$ & $    57.5$ & $100$ & $    60.2$ & $100$ & $    59.7$\\
	\end{tabular}
\end{ruledtabular}
\end{table}

\begin{figure}[htb]
	\centering
	\includegraphics[width=8.5cm]{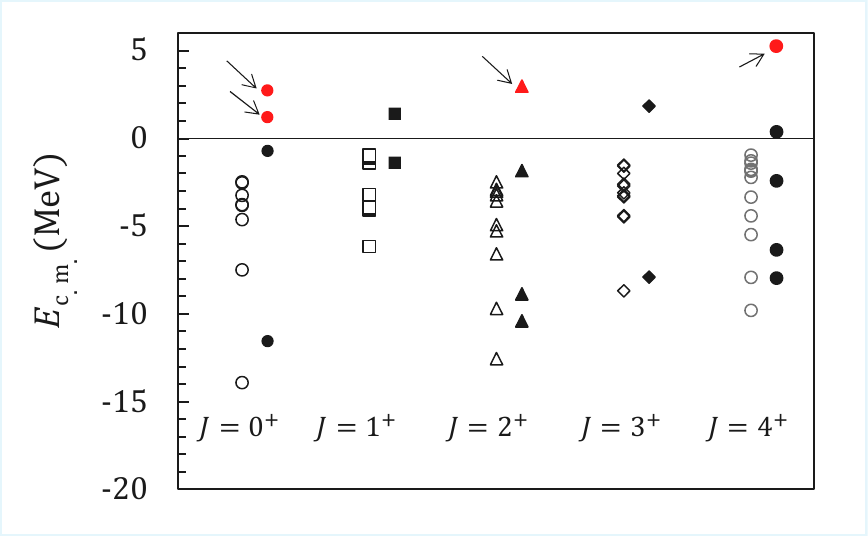}
	\caption{$^{24}$Mg states in the multichannel calculation (filled symbols) compared to experiment (open symbols). The red marks (also indicated by arrows) correspond to the resonances present in the GCM fusion cross section (see Sec.~\ref{sec6}).}
	\label{fig_24mg2}
\end{figure}

In Fig.~\ref{fig_24mg2}, I present bound states and resonances. All $\carb$ and $\neon$ states shown in Fig.~\ref{fig_spec1} are included in the calculation. As mentioned before, the lowest of $0^+,2^+$ and $4^+$ states are slightly too high ($\approx 2$ MeV) which is due to the lack of the $p + ^{23}$Na and $n + ^{23}$Mg partitions. Of course the experimental level density is high, and a two-cluster model, even with many channels, cannot be expected to reproduce all $^{24}$Mg states.
In Fig.~\ref{fig_24mg2}, I indicate by arrows resonances which play a role in the low-energy fusion cross section (see Sec.~5.4). Two $0^+$, one $2^+$ and two $4^+$ resonances are found above the $\cc$ threshold.

In microscopic single-channel calculations \cite{BP82,BD84}, ``molecular states" were found. These states are expected to have a dominant 	$\text{C}(0^+_1)+\text{C}(0^+_1)$ structure.  In the multichannel calculation, however, these resonances do not show up. The wave functions of $^{24}$Mg bound states and resonances are spread over several channels. This is supported by the semi-microscopic calculation of Suzuki and Hecht \cite{SH82} who investigate $\cc$ resonances by including the $\cc$ and $\ane$ partitions in their ground state.

\subsection{$\cc$ elastic scattering}
The present model, using the same calculation conditions, can be used to study elastic scattering. This process is not directly related to fusion, but is based on the same scattering wave functions. Elastic scattering is therefore an excellent test of the model. 

\begin{figure}[htb]
	\centering
	\includegraphics[width=7.6cm]{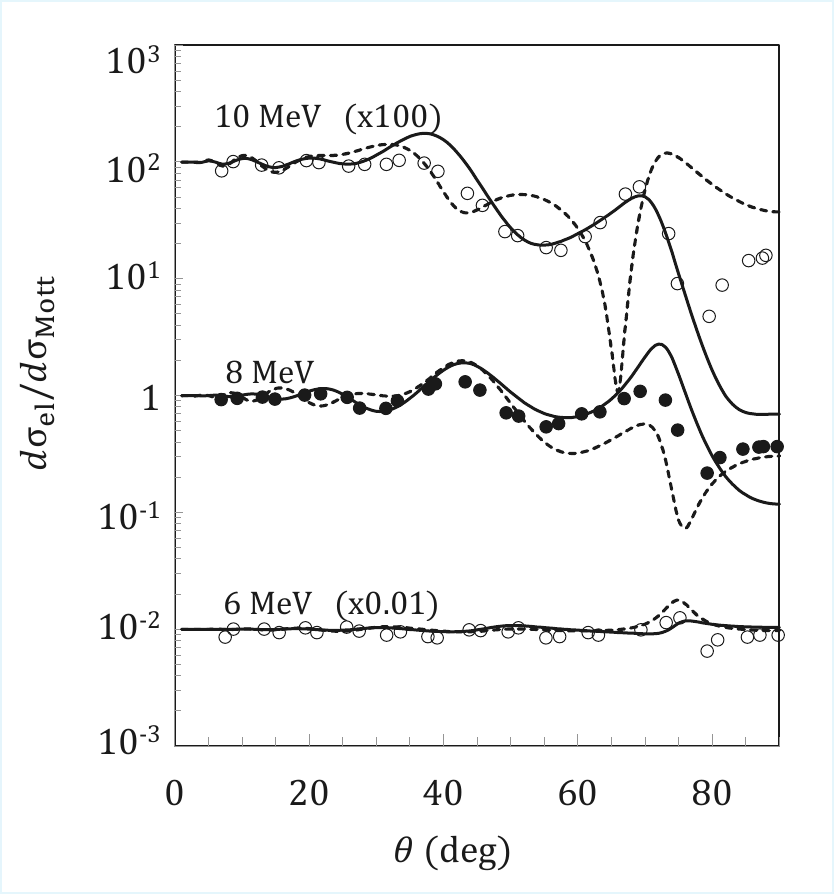}
	\caption{Ratios of the elastic and Mott cross sections at 3 c.m.\ energies around the Coulomb barrier. The solid lines correspond to the multichannel calculation, and the dotted lines to the single-channel approximation. The experimental data are taken from Ref.~\cite{TFG80}.}
	\label{fig_elas}
\end{figure}

I present elastic cross sections (divided by the Mott cross section) in Fig.~\ref{fig_elas}, and compare them to experiment \cite{TFG80}. Below the Coulomb barrier, and even more at stellar energies, this ratio is close to unity. Consequently, I have selected c.m.\ energies slightly above the Coulomb barrier (6, 8,  and 10 MeV). I compare the single-channel calculation (dotted lines) with the multichannel model (solid lines). At $6$ MeV, the nuclear effects are minimal, and the cross section weakly depends on the angle. When the energy increases (up to $10$ MeV), however, the cross section significantly deviates from the Mott cross section. 

The multichannel calculation accurately reproduces the oscillations observed in the experimental data. Clearly, the single-channel approximation is rather far from experiment, and this disagreement increases with energy. The excellent results obtained with the multichannel calculation suggests that absorption effects are well simulated by $\cc$ and $\ane$ channels, even if many channels are closed (see Fig.~\ref{fig_spec1}). These virtual off-shell couplings obviously help to improve the description of $\cc$ elastic scattering. The application of this model to the fusion process is therefore expected to be reliable. 
	
\section{Fusion cross section}
\label{sec6}

\begin{table*}[!]
	\caption{Resonance properties (energies, partial widths and amplitudes ${\mathcal N}_{\gamma}$).}
	\label{table3}
	\begin{ruledtabular}
		\begin{tabular}{lccccc}
			& $4^+$ & $0^+$ & $0^+$ & $2^+$ & $4^+$ \\
			\hline
			$E_{\rm c.m.}$ (MeV) & 0.30 & 1.14 & 2.6 & 2.9 & 5.1 \\ \\
			$\Gamma$ (MeV)\\
			\hline
			$\cc$ & $7.1\times 10 ^{-57}$ & $1.2\times 10 ^{-21}$ & $3.9\times 10 ^{-9}$ & $1.5\times 10 ^{-8}$ & $1.3\times 10 ^{-4}$ \\
			$\ane(0^+)$ & $8.5\times 10 ^{-7}$ & $2.6\times 10^{-2}$ & 0.33 & 0.13 & 0.25 \\
			$\ane(2^+)$ & $3.8 \times 10^{-3}$ & 0.25 & 0.58 & 0.20 & 0.44 \\ \\
			${\mathcal N}_{\gamma}$ \\
			\hline
			$\cc$ & 0.16 & 0.04 & 0.20 & 0.22 & 0.16 \\
			$\ane(0^+)$ & 0.05 & 0.03 & 0.07 & 0.04 & 0.05 \\
			$\ane(2^+)$ & 0.11 & 0.30 & 0.19 & 0.12 & 0.17 
		\end{tabular}
	\end{ruledtabular}
\end{table*}

\begin{figure}[htb]
	\centering
	\includegraphics[width=8.6cm]{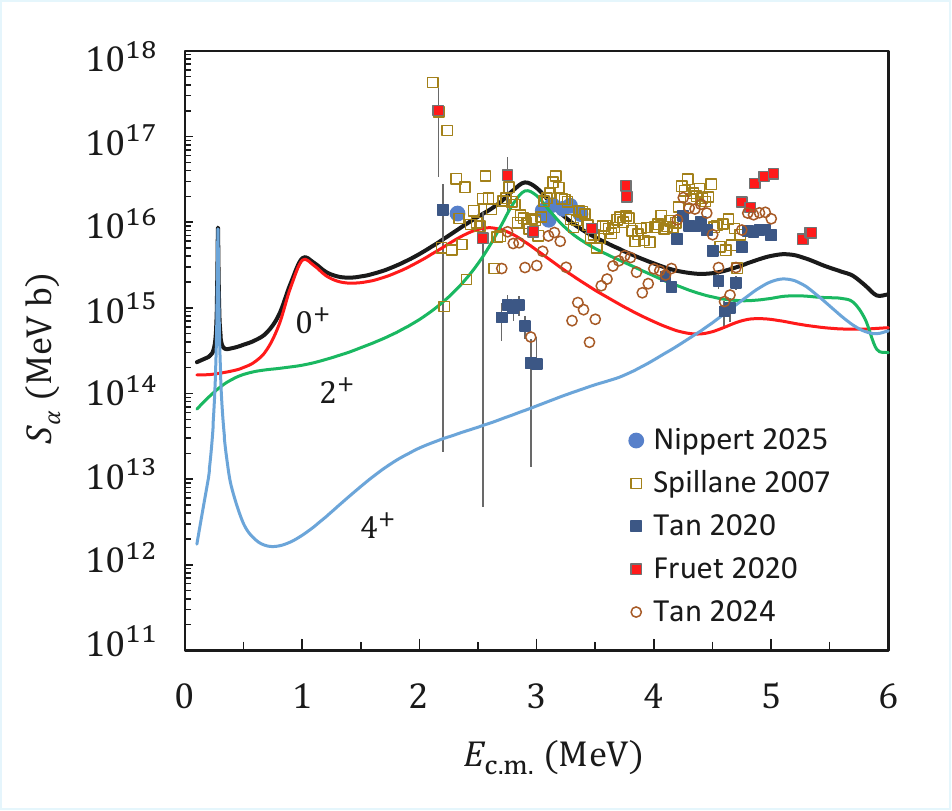}
	\caption{Fusion $S$-factor in the $\alpha$ channel. The experimental data are taken from Refs.~\cite{SRR07,TBD20,FCH20,TGL24,NCH25}.}
	\label{fig_fusion}
\end{figure}

In Fig.~\ref{fig_fusion}, I present the fusion cross section \eqref{eq6} converted to the $S$-factor
\begin{align}
S(E) = \sigma(E) E \exp(2\pi \eta + gE),
\label{eq16}
\end{align}
where $\eta$ is the Sommerfeld parameter, and where coefficient $g=0.46$ MeV$^{-1}$ is usually introduced to compensate the energy dependence of the cross section. I have selected experimental data considering the $\alpha$ channel only, since the proton and neutron channels are not included in my calculation. The cross section is decomposed in partial waves, in order to underline the role of the resonances. The model predicts $0^+$ and $2^+$ states around 3 MeV. At low energies, narrow $2^+$ and $4^+$ resonances are present in the calculation.

The resonances observed in the fusion cross section are consistent with the $^{24}\mathrm{Mg}$ states shown in Fig.~\ref{fig_24mg2}. They are indicated by arrows in that figure. The properties of these resonances are detailed in Table \ref{table3}. As they are located below the Coulomb barrier ($V_C \approx 6$ MeV), the $\cc$ partial widths are extremely small. The main contribution to the widths comes from the $\ane$ channels. Since the widths are strongly correlated to the energies, the structure of the resonances is better illustrated by the overlaps \eqref{eq13}, which do not depend on energy. In most cases, there is no dominant channel. The structure of the resonances is a mixture of several configurations.

In the literature, there is a debate about the possible fusion hindrance in the $\cc$ reaction at low energies \cite{JRB07}. Some calculations predict a steep decrease of the $S$ factor at energies well below the Coulomb barrier (see the discussion in Ref.~\cite{UH25}). This effect is expected below $E \lesssim 2$ MeV, but the cross section is too small to be measured in the laboratory, and no definite conclusion can be drawn. 

In optical-model calculations, hindrance effects are analyzed through the function
\begin{align}
	L(E) = \frac{d\ln (\sigma E)}{dE}
	\label{eq17}
\end{align}
which is parameterized by 
\begin{align}
	L(E) \simeq A_0+B_0/E^n,
	\label{eq18}
\end{align}
with $n=1.5$ (this choice, however is questioned in Ref.~\cite{UH25}). Coefficients $A_0$ and $B_0$ are fitted, and depend on the potential. In my calculation, the existence of resonances makes the parametrization \eqref{eq18} non applicable. However, the GCM $S$-factor is consistent with fusion hindrance, as a decrease of the $S$-factor is found at low energies. Of course this conclusion is limited to the $\alpha$ channel, and could be affected by the proton and neutron channels. Calculations with more configurations are necessary before drawing definite conclusions about fusion hindrance.

\section{Conclusion and outlook}
\label{sec7}
The present work paves the way for a fully microscopic approach to fusion reactions. The $\cc$ reaction is extremely important in nuclear astrophysics, and raises several open issues, such as the presence of resonances or the fusion hindrance below the Coulomb barrier. A microscopic theory, with an explicit treatment of the continuum and of the various reaction channels, is a numerical challenge. However, this approach provides a consistent description of fusion, elastic and inelastic scattering and of the $^{24}$Mg spectroscopy. All physical quantities are based on a nucleon-nucleon interaction only. Data on elastic scattering are well described, and I have shown that the multichannel model strongly improves the single-channel calculation.

My RGM-GCM calculation goes far beyond previous $\cc$ microscopic calculations, that used a single-channel approximation \cite{BP82,SH82,BD84}. I include several $\carb$ states, as well as $\ane$ channels. One of the consequences is that the so-called ``molecular states" found in single-channel calculation, are not present. These states are expected to have a dominant $\cc$ structure. However, my analysis of $^{24}$Mg bound states and resonances shows that the wave functions are spread over many channels. In the multichannel calculation, I do not find bound states or resonances with a dominant	$\text{C}(0^+_1)+\text{C}(0^+_1)$ configuration.

The fusion cross section is consistent with experiment, even though experimental uncertainties remain, even around the Coulomb barrier. Extrapolating the data down to stellar energies is a delicate problem. I predict resonances, in particular $0^+$ and $2^+$ broad states near the Coulomb barrier. Additionally, narrow $0^+$ and $4^+$ resonances are present in the GCM cross section. The model could be improved by including the proton and neutron channels, which are neglected in this exploratory study. Introducing negative-parity states of $^{20}$Ne could also play a role. The inclusion of these channels represents a challenge for future calculations. 

\section*{Acknowledgments}
This work was supported by the Fonds de la Recherche Scientifique - FNRS under Grant Numbers 4.45.10.08 and J.0065.22. It benefited from computational resources made available on the Tier-1 supercomputer of the 
F\'ed\'eration Wallonie-Bruxelles, infrastructure funded by the Walloon Region under the grant agreement No. 1117545.

%\bibliography{../biblio.bib}
%\bibliographystyle{apsrev4-2}


\end{document}